\documentclass[aps,prd,preprintnumbers,superscriptaddress,nofootinbib,showpacs,twocolumn]{revtex4}%
\usepackage[dvipdfmx]{graphicx}
\usepackage{bm,latexsym,amsmath,amssymb,amsfonts,mathrsfs}
\usepackage{color}
\input{colordvi.tex}
\usepackage[dvipdfmx]{hyperref}
\usepackage{url}
\hypersetup{
    colorlinks=true,
    citecolor=cyan,
}
\newcommand*{\D}{{\rm d}}

\begin{document}

\title{Generic instabilities of non-singular cosmologies in Horndeski theory:\\
a no-go theorem}

\author{Tsutomu~Kobayashi}
\email[Email: ]{tsutomu"at"rikkyo.ac.jp}
\affiliation{Department of Physics, Rikkyo University, Toshima, Tokyo 171-8501, Japan
}

\begin{abstract}
The null energy condition can be violated stably
in generalized Galileon theories, which gives rise to
the possibilities of healthy non-singular cosmologies.
However, it has been reported that
in many cases cosmological solutions are plagued with instabilities
or have some pathologies
somewhere in the whole history of the universe.
Recently, this was shown to be generically true
in a certain subclass of the Horndeski theory.
In this short paper, we extend this no-go argument to the full Horndeski theory
and show that non-singular models (with flat spatial sections)
in general suffer from either gradient instabilities
or some kind of pathology in the tensor sector.
This implies that one must go beyond the Horndeski theory to implement healthy non-singular cosmologies.
\end{abstract}

\preprint{RUP-16-19}
\maketitle
%%%%%%%%%%%%%%%%%%%%%%%%%%%%%%%%%%%%%%%%%%%%%%%%%%%%%%%%%%%%%%%%%%%%%%

\section{Introduction}

Inflation~\cite{Guth:1980zm,Starobinsky:1980te,Sato:1980yn}
is now the strongest candidate of the early universe scenario
that explains current cosmological observations consistently.
Nonetheless, alternative scenarios deserve to be considered as well.
First, in order to be convinced that inflation indeed occurred in the early stage of the universe,
all other possibilities must be ruled out. Second, even inflation
cannot resolve the problem of the initial singularity~\cite{Borde:1996pt}.
It is therefore well motivated to study how good and how bad
alternative possibilities are compared to inflation.
Non-singular stages in the early universe, such as
contracting and bouncing phases~\cite{Battefeld:2014uga},
cannot only be something that replaces inflation,
but also ``early-time'' completion of inflation just to get rid of the initial singularity.
In this paper, we address whether healthy non-singular cosmologies
can be implemented in the framework of general scalar-tensor theories.

If gravity is described by general relativity and
the energy-momentum tensor $T_{\mu\nu}$ of matter satisfies the null energy condition (NEC),
that is, $T_{\mu\nu} k^\mu k^\nu\ge 0$ for every null vector $k^\mu$,
then (assuming flat spatial sections)
it follows from the Einstein equations that
$\D H/\D t \le 0$, where $H$ is the Hubble parameter.
This implies that an expanding universe yields a singularity in the past, while
NEC violation could lead to singularity-free cosmology.
However, violating the NEC in a healthy manner turns out to be challenging.
The NEC is by construction satisfied for a canonical scalar field,
$T_{\mu\nu} k^\mu k^\nu=\dot\phi^2\ge 0$.
In a general non-canonical scalar-field theory whose Lagrangian is dependent on
$\phi$ and its first derivative~\cite{ArmendarizPicon:1999rj,Garriga:1999vw},
the NEC can be violated, but
NEC-violating cosmological solutions are unstable
because the curvature perturbation has the wrong sign kinetic term.

Galileon theory~\cite{Nicolis:2008in} and its generalizations~\cite{Deffayet:2009wt,Deffayet:2011gz}
involve the scalar field whose Lagrangian contains second derivatives of $\phi$
while maintaining the second-order nature of the equation of motion
and thus erasing the Ostrogradsky instability. In contrast to the previous case,
it was found that the NEC and the stability of cosmological solutions are uncorrelated
in Galileon-type theories~\cite{Creminelli:2010ba}.
This fact gives rise to healthy NEC-violating models of
Galilean genesis~\cite{Creminelli:2010ba,Creminelli:2012my,Hinterbichler:2012fr,Hinterbichler:2012yn,Easson:2013bda,Nishi:2015pta,Nishi:2016wty}
and stable non-singular bouncing solutions~\cite{Qiu:2011cy,Easson:2011zy,Osipov:2013ssa},
as well as novel dark energy and inflation models with interesting phenomenology~\cite{Deffayet:2010qz,Kobayashi:2010cm}.
See also a recent review~\cite{Rubakov:2014jja}.

Although the Galileon-type theories do admit a stable early stage without an initial singularity,
the genesis/bouncing universe must be interpolated to a subsequent (possibly conventional) stage
and the stable early stage does not mean that the cosmological solution is stable at all times
during the whole history.
Several explicit
examples~\cite{Cai:2012va,Koehn:2013upa,Battarra:2014tga,Qiu:2015nha,Wan:2015hya,Pirtskhalava:2014esa,Kobayashi:2015gga}
show that
the sound speed squared of the curvature perturbation becomes negative
at around the transition between the genesis/bouncing phase and the subsequent phase,
leading to gradient instabilities. In some cases the universe can experience a healthy bounce,
but then the solution has some kind of singularity in the past or future~\cite{Easson:2011zy}.
Although the gradient instabilities can be cured by introducing higher spatial derivative
terms~\cite{Pirtskhalava:2014esa,Kobayashi:2015gga}
and there are some models in which the strong coupling scale cuts off the instabilities~\cite{Koehn:2015vvy},
it would be preferable if
the potential danger could be removed from the beginning.
The next question to ask therefore is whether the appearance of instabilities is generic or
a model-dependent nature.
For general dilation invariant theories a no-go theorem was given in Ref.~\cite{Rubakov:2013kaa}.
(A counterexample was presented in Ref.~\cite{Elder:2013gya}, but it has an initial singularity.)
Recently, it was clearly shown in Ref.~\cite{Libanov:2016kfc} that bouncing and genesis models suffer from
instabilities or have singularities for the scalar-tensor theory whose Lagrangian is of the form
\begin{align}
&{\cal L}=\frac{R}{2\kappa}+G_2(\phi, X)-G_3(\phi, X)\Box\phi,
\notag \\
&X:=-\frac{1}{2}g^{\mu\nu}\partial_\mu\phi\partial_\nu\phi,\label{KGB}
\end{align}
where $R$ is the Ricci scalar.
This Lagrangian is widely used in the attempt to obtain non-singular stable cosmology.

The Lagrangian~(\ref{KGB}) forms a subclass of the most general scalar-tensor theory
with second-order field equations, i.e., the Horndeski theory~\cite{Horndeski:1974wa}.
The goal of this short paper is to generalize the no-go argument of Ref.~\cite{Libanov:2016kfc}
to the {\em full} Horndeski theory.

\section{No-go theorem}

We consider the Horndeski theory~\cite{Horndeski:1974wa} in its complete form,
\begin{align}
S=\int\D^4 x\sqrt{-g}{\cal L}_H,
\end{align}
where
\begin{align}
{\cal L}_H&=G_2(\phi, X)-G_3(\phi, X)\Box\phi
\notag \\ & \quad
+G_4(\phi, X)R
+G_{4,X}\left[(\Box\phi)^2-(\nabla_\mu\nabla_\nu\phi)^2\right]
\notag \\ & \quad
+G_5(\phi, X)G^{\mu\nu}\nabla_\mu\nabla_\nu\phi
-\frac{1}{6}G_{5,X}\bigl[(\Box\phi)^3
\notag \\ & \quad
-3\Box\phi(\nabla_\mu\nabla_\nu\phi)^2
+2(\nabla_\mu\nabla_\nu\phi)^3\bigr].
\label{Hor_L}
\end{align}
(The Lagrangian here is written in the form of the generalized Galileon~\cite{Deffayet:2011gz},
but the two theories are in fact equivalent~\cite{Kobayashi:2011nu}.)
In the full Horndeski theory, we have four arbitrary functions of
the scalar field $\phi$
and $X=-g^{\mu\nu}\partial_\mu\phi\partial_\nu\phi/2$.
The scalar field is coupled to the Ricci scalar $R$ and the Einstein tensor $G_{\mu\nu}$
in the particular way shown above. The structure of the Lagrangian~(\ref{Hor_L})
guarantees the second-order nature of the field equations.

The equations of motion governing the background cosmological evolution
can be obtained by substituting
$\D s^2=-N^2(t)\D t^2+a^2(t)\delta_{ij}\D x^i\D x^j$ and $\phi=\phi(t)$
to the Horndeski action
and varying it with respect to $N$, $a$, and $\phi$~\cite{Kobayashi:2011nu}.
In this paper, we only consider a spatially flat universe.

Linear perturbations around a spatially flat FLRW spacetime in the Horndeski theory were studied in
Ref.~\cite{Kobayashi:2011nu}. Taking the unitary gauge, $\delta\phi=0$,
the spatial part of the metric can be written as $\gamma_{ij}=a^2(t)e^{2\zeta}\left(e^h\right)_{ij}$,
where $\zeta$ is the curvature perturbation and $h_{ij}$ is the tensor perturbation.
The quadratic actions for $h_{ij}$ and $\zeta$ are given, respectively, by~\cite{Kobayashi:2011nu}
\begin{align}
S_h^{(2)}=\frac{1}{8}\int\D t\D^3x \,a^3\left[{\cal G}_T\dot h_{ij}^2
-\frac{{\cal F}_T}{a^2}(\partial h_{ij})^2\right],
\end{align}
and
\begin{align}
S_\zeta^{(2)}=\int\D t\D^3x \,a^3\left[{\cal G}_S\dot \zeta^2
-\frac{{\cal F}_S}{a^2}(\partial \zeta)^2\right].
\end{align}
Here, the coefficients are written as
\begin{align}
{\cal F}_T&:=2 \left[ G_4-X \left(\ddot\phi G_{5,X}+G_{5,\phi}\right)\right],
\\
{\cal G}_T&:=2 \left[ G_4-2XG_{4,X}-X \left(H\dot\phi G_{5,X}-G_{5,\phi}\right)\right],
\end{align}
where a dot denotes differentiation with respect to
cosmic time $t$,
while ${\cal F}_S$ and ${\cal G}_S$
have more complicated dependence on the functions $G_2$, $G_3$, $G_4$, and $G_5$,
the explicit forms of which are found in Ref.~\cite{Kobayashi:2011nu}.
It is reasonable to assume that ${\cal F}_T$, ${\cal G}_T$, ${\cal F}_S$, and ${\cal G}_S$ are
smooth functions of time.
To avoid ghost and gradient instabilities, we require that
\begin{align}
{\cal F}_T> 0,\quad{\cal G}_T>0,\quad
{\cal F}_S> 0,\quad{\cal G}_S>0.
\end{align}
If $\phi$ is minimally coupled to gravity, we have $G_4=\;$const and $G_5=0$,
and hence ${\cal F}_T={\cal G}_T=\;$const.
In other words, the time evolution of ${\cal F}_T$ and ${\cal G}_T$
is caused by non-minimal coupling to gravity.

The crucial point for the no-go argument is that
${\cal F}_S$ is generically of the form
\begin{align}
{\cal F}_S=\frac{1}{a}\frac{\D \xi}{\D t}-{\cal F}_T,
\end{align}
where
\begin{align}
\xi:=\frac{a{\cal G}_T^2}{\Theta},
\end{align}
with
\begin{align}
\Theta
& := -\dot\phi XG_{3,X}+2HG_4-8HXG_{4,X}-8HX^2G_{4,XX}
\notag \\ & \quad
+\dot\phi G_{4,\phi}+2X\dot\phi G_{4,\phi X}
+2HX\left(3G_{5,\phi}+2XG_{5,\phi X}\right)
\notag \\ & \quad
-H^2\dot\phi\left(5X G_{5,X}+2X^2G_{5,XX}\right).
\end{align}
Since $\Theta$ is something written in terms of $\phi$ and $H$,
it is supposed to be a smooth function of time which is finite everywhere.
This then implies that $\xi$ can never vanish except at a singularity, $a=0$.
The absence of gradient instabilities is equivalent to
\begin{align}
\frac{\D\xi}{\D t}> a{\cal F}_T> 0.
\label{st2}
\end{align}
Integrating Eq.~(\ref{st2}) from some $t_{\rm i}$ to $t_{\rm f}$, we obtain
\begin{align}
\xi_{\rm f}-\xi_{\rm i}> \int_{t_{\rm i}}^{t_{\rm f}}a{\cal F}_T\D t.\label{key1}
\end{align}
This is the key equation for the following argument,
and it was used to prove the no-go theorem in
the subclass of the Horndeski theory with $G_4=\;$const and $G_5=0$ in Ref.~\cite{Libanov:2016kfc}.
Remarkably, it turns out that essentially the same equation holds in the full Horndeski theory.

Now, consider a non-singular universe which
satisfies $a>\;$const ($>0$) for $t\to -\infty$ and
is expanding for large $t$.
The integral in the right hand side of Eq.~(\ref{key1})
may be convergent or not as one takes $t_{\rm f}\to\infty$ and $t_{\rm i}\to-\infty$,
depending on the asymptotic behavior of ${\cal F}_T$.
To allow the integral to converge, it is necessary that ${\cal F}_T$
approaches zero sufficiently fast in the asymptotic past or future.
For the moment let us focus on the case where the integral is not convergent.

Suppose that $\xi_{\rm i}<0$.
Equation~(\ref{key1}) reads
\begin{align}
-\xi_{\rm f}<|\xi_i|-\int_{t_{\rm i}}^{t_{\rm f}}a{\cal F}_T\D t.
\end{align}
Since the integral is an increasing function of $t_{\rm f}$,
the right hand side becomes negative for sufficiently large $t_{\rm f}$.
We therefore have $\xi_{\rm f}>0$, which means that $\xi$ crosses zero.\footnote{We do not
allow for discontinuity in $\xi$ because ${\cal F}_S$ is supposed to be smooth.
(This means that $\Theta$ cannot cross zero.)}
This is never possible in a non-singular universe.
It is therefore necessary to have $\xi>0$ everywhere.
Writing Eq.~(\ref{key1}) as
\begin{align}
-\xi_{\rm i}>-\xi_{\rm f}+\int_{t_{\rm i}}^{t_{\rm f}}a{\cal F}_T\D t,\label{key2}
\end{align}
we see that the right hand side will be positive for $t_{\rm i}\to -\infty$
and hence $\xi_{\rm i}<0$.
However, this is in contradiction to the assumption that $\xi$ is always positive.
Thus, we have generalized the no-go argument of Ref.~\cite{Libanov:2016kfc}
to the full Horndeski theory.

The same no-go theorem holds even in the presence of another field,
provided at least that the field is described by
\begin{align}
{\cal L}_\chi=P(\chi, Y),\quad Y:=-\frac{1}{2}g^{\mu\nu}\partial_\mu\chi\partial_\nu\chi,
\end{align}
which is not coupled to the Horndeski field $\phi$ directly.

Now there are two degrees of freedom in the scalar sector of cosmological perturbations.
In terms of
\begin{align}
\Vec{y}:=\left(\zeta,\, \frac{\Theta}{{\cal G}_T}\frac{\delta\chi}{\dot\chi}\right),
\end{align}
the quadratic action can be written in the form~\cite{DeFelice:2010gb,DeFelice:2011bh,Kobayashi:2013ina}
\begin{align}
S^{(2)}&=\int\D t\D^3x a^3\left[\dot{\Vec{y}}\,\mathbf{G}\,\dot{\Vec{y}}-\frac{1}{a^2}\partial \Vec{y}\,
{\mathbf F}\,\partial\Vec{y}+\cdots\right],
\end{align}
where
\begin{eqnarray}
  \mathbf{G} = \left(
    \begin{array}{cc}
      {\cal G}_S+Z & -Z \\
      -Z & Z
    \end{array}
  \right), 
\quad
  \mathbf{F} = \left(
    \begin{array}{cc}
      {\cal F}_S & -c_s^2Z \\
      -c_s^2Z & c_s^2Z
    \end{array}
  \right), 
\end{eqnarray}
with
\begin{align}
c_s^2:=\frac{P_{,Y}}{P_{,Y}+2YP_{,YY}},
\quad
Z:=\left(\frac{{\cal G}_T}{\Theta}\right)^2\frac{YP_{,Y}}{c_s^2}.
\end{align}
Here, ${\cal G}_S$ and ${\cal F}_S$ were defined previously and $c_s$ is the sound speed of the $\chi$ field.
We have the relation $2YP_{,Y}=\rho+P$, where
$\rho$ is the energy density of $\chi$ and $P$ corresponds to the pressure of $\chi$.

Ghost instabilities can be evaded if $\mathbf{G}$ is a positive definite matrix.
The condition amounts to
\begin{align}
{\cal G}_S>0,\quad \frac{YP_{,Y}}{c_s^2}>0.
\end{align}
The propagation speed $v$ can be determined by solving
\begin{align}
{\rm det}(v^2\mathbf{G}-\mathbf{F})=0,
\end{align}
yielding the condition for the absence of gradient instabilities,
\begin{align}
c_s^2> 0,\quad \frac{{\cal F}_S-c_s^2Z}{{\cal G}_S}> 0.
\end{align}
We thus have the inequality
\begin{align}
{\cal F}_S> \frac{1}{2}\left(\frac{{\cal G}_T}{\Theta}\right)^2\left(\rho+P\right)>0,
\end{align}
and taking the same way we can show the no-go theorem
for the Horndeski $+$ k-essence (or a perfect fluid) system.

The no-go theorem we have thus established can be circumvented
only if ${\cal F}_T$ approaches zero sufficiently fast
either in the asymptotic past or the future,
given the assumption that the evolution of the scale factor is non-singular.\footnote{The ``modified
genesis'' model proposed in Ref.~\cite{Libanov:2016kfc} evades the no-go theorem by the use of
the vanishing scale factor in the asymptotic past.
In contrast, we are assuming that the expansion history is non-singular everywhere, i.e., $a\ge\;$const.}
The normalization of vacuum quantum fluctuations tells us that they would grow and diverge as ${\cal F}_T\to 0$,
and hence the tensor sector is pathological in the asymptotic past or future.\footnote{One could resolve this issue
by the particular, fine-tuned evolution of ${\cal G}_T$, which would offer a loophole.}
In the next section, we will demonstrate that, in contrast to the cases
in Refs.~\cite{Pirtskhalava:2014esa,Kobayashi:2015gga},
it is indeed possible to construct a model that exhibits
a stable transition from the Galilean genesis to inflation
by allowing for some kind of pathology in the tensor sector due to vanishing ${\cal F}_T$.

\section{Stable transition from genesis to de Sitter with pathologies in the past}

Let us turn to study a specific setup as an example:
Galilean genesis followed by inflation.
Such an expansion history was proposed in Refs.~\cite{Pirtskhalava:2014esa,Kobayashi:2015gga}
as early-time completion of the inflationary universe,
and there it was pointed out that the sound speed squared (or more specifically ${\cal F}_S$)
becomes negative at the transition from the genesis phase to inflation.
This is consistent with the no-go theorem, because
in the genesis phase we have $a\to\;$const as $t\to-\infty$ and ${\cal F}_T=\;$const.
The resultant gradient instability is cured by the introduction of
higher order spatial derivatives arising in the effective field theory approach~\cite{Pirtskhalava:2014esa}
or in theories beyond Horndeski~\cite{Gleyzes:2014dya,Gao:2014soa,Kobayashi:2015gga}.

Working within the second-order theory, i.e., the Horndeski theory,
we are going to show in this section that the stable transition is indeed possible
if ${\cal F}_T\to 0$ as $t\to-\infty$ so that the integral in Eq.~(\ref{key1}) is convergent.
To do so it is more convenient to use the ADM form of the action
rather than the original covariant one~\cite{Kobayashi:2015gga}.
The ADM decomposition of the Horndeski Lagrangian leads to~\cite{Gleyzes:2014dya}
\begin{align}
{\cal L}&=A_2(t,N)+A_3(t,N)K+A_4(t,N)\left(K^2-K_{ij}^2\right)
\notag \\ &\quad +A_5(t,N)\left(K^3-3KK_{ij}^2+2K_{ij}^3\right)
\notag \\ &\quad +B_4(t,N)R^{(3)}+B_5(t,N)K^{ij}G_{ij}^{(3)},\label{LADM}
\end{align}
where $\phi=\;$const hypersurfaces are taken to be constant time hypersurfaces,
and $K_{ij}$, $R_{ij}^{(3)}$, and $G_{ij}^{(3)}$ are the
extrinsic curvature, the Ricci tensor, and the Einstein tensor of the
spatial slices.
The functions of $\phi$ and $X$ in the covariant Lagrangian are
now the functions of $t$ and the lapse function $N$.
Two of the six functions in the ADM Lagrangian~(\ref{LADM})
are subject to the constraints
\begin{align}
A_4=-B_4-N\frac{\partial B_4}{\partial N},\quad A_5=\frac{N}{6}\frac{\partial B_5}{\partial N},
\end{align}
in accordance with the fact that
there are four arbitrary functions in the Horndeski theory.

The specific example we are going to study
is given by the functions of the form
\begin{align}
&A_2= f^{-2(\alpha+1)-\delta}a_2(N),
\quad
&&A_3= f^{-2\alpha-1-\delta}a_3(N),
\notag \\
&A_4=-B_4=- f^{-2\alpha},
\quad
&&A_5=B_5=0,\label{Aform}
\end{align}
where $f=f(t)$ is dependent only on $t$,
and $\alpha$ and $\delta$ are constant parameters
satisfying $2\alpha>1+ \delta>1$.
This class of models is similar to but different from that in Ref.~\cite{Kobayashi:2015gga}.
The covariant form of the Lagrangian can be recovered by re-introducing the scalar field, e.g.,
through $-t=e^{-\phi}$ and $N^{-1}=e^{-\phi}\sqrt{2X}$ and
using the Gauss-Codazzi equations.
In terms of $G_2(\phi, X)$, $G_3(\phi,X)$, ..., the Lagrangian is written in
a slightly more complicated form~\cite{NKinprep}.
Without moving to the covariant description,
one can derive the equations of motion for the homogeneous background
directly from variation of the ADM action with respect to $N$ and the scale factor $a$.

%----------------------------------------------------------------%
\begin{figure}[tbp]
  \begin{center}
  \includegraphics[keepaspectratio=true,width=80mm]{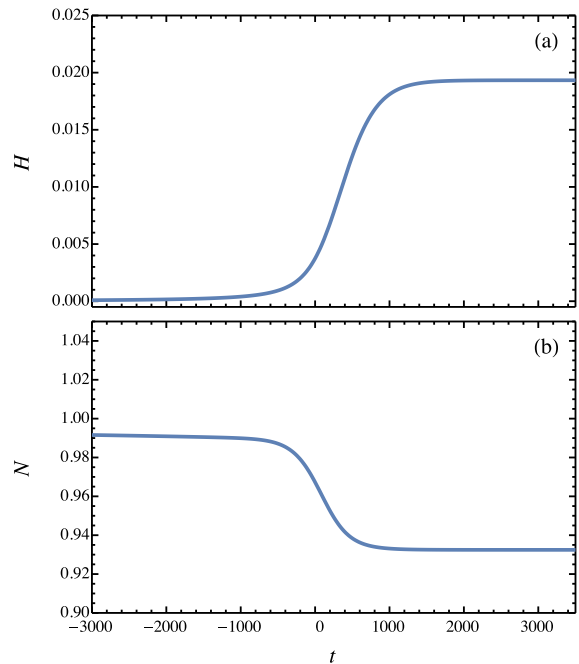}
  \end{center}
  \caption{Evolution of the Hubble parameter and the lapse function around the genesis--de Sitter transition.}%
  \label{fig:hubble.eps}
\end{figure}
%----------------------------------------------------------------%
\begin{figure}[tbp]
  \begin{center}
  \includegraphics[keepaspectratio=true,width=80mm]{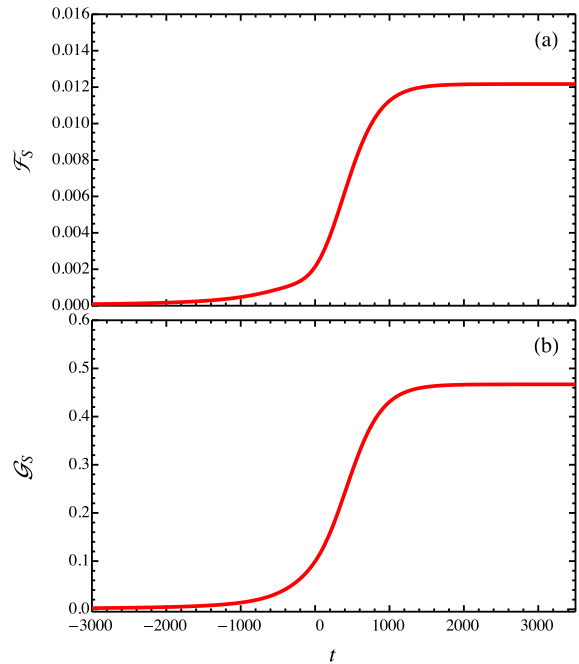}
  \end{center}
  \caption{${\cal F}_S$ and ${\cal G}_S$ around the genesis--de Sitter transition.}%
  \label{fig:stable.eps}
\end{figure}
%----------------------------------------------------------------%

The evolution of the Hubble parameter, $H:=N^{-1}\D \ln a/\D t$,
is dependent crucially on the choice of $f(t)$, and to describe the genesis to de Sitter transition
we take $f(t)$ such that $f\sim c(-t)\gg 1$ ($c>0$) in the past and $f\sim\;$const in the future.
In the early time, we have an approximate solution of the form
\begin{align}
H\simeq \frac{{\rm const}}{(-t)^{1+\delta}},
\end{align}
and hence the universe starts expanding from Minkowski,
\begin{align}
a\simeq 1+\frac{{\rm const}}{(-t)^\delta},
\end{align}
with $N\simeq\;$const. In the late time where $f\simeq\;$const, we have
an inflationary solution $H\simeq\;$const, again with $N\simeq\;$const.
For all the models described by~(\ref{Aform}), we have
\begin{align}
{\cal F}_T={\cal G}_T = f^{-2\alpha}>0,
\end{align}
and hence the stability conditions for the tensor modes are fulfilled.
Since $a{\cal F}_T\sim (-t)^{-2\alpha}$ with $2\alpha >1$ as $t\to -\infty$,
${\cal F}_T$ possesses the desired property to evade the no-go theorem.

As a concrete example, we consider
\begin{align}
a_2=-\frac{1}{N^2}+\frac{1}{3N^4},\quad a_3=\frac{1}{4N^3},
\end{align}
with $\alpha=1$, $\delta=1/2$,
and
\begin{eqnarray}
f(t)=\frac{c}{2}\left[-t+\frac{\ln(2\cosh(st))}{s}\right]+f_{1},\label{f:ex1}
\end{eqnarray}
where the parameters are taken to be
$c=10^{-1}$, $f_{1}=10$, and $s=2\times 10^{-3}$.
The background equations are solved numerically to give the
evolution of $H$ and $N$ as shown in Fig.~\ref{fig:hubble.eps}.
It can be seen that the universe indeed undergoes the genesis phase followed by inflation.
For this background solution we evaluate ${\cal F}_S$ and ${\cal G}_S$ numerically
to judge its stability. As presented in Fig.~\ref{fig:stable.eps}, we find that
${\cal F}_S$ and ${\cal G}_S$ remain positive in the whole expansion history.
This is in contrast to the similar example in Refs.~\cite{Pirtskhalava:2014esa,Kobayashi:2015gga}
which has ${\cal F}_S<0$ around the transition.

Although the present model can circumvent the gradient instability at the
genesis--de Sitter transition, some pathologies arise in the $t\to-\infty$ limit.
We see that ${\cal F}_T,\, {\cal G}_T\sim (-t)^{-2\alpha}$
and
${\cal F}_S,\,{\cal G}_S\sim (-t)^{-2\alpha+\delta}$
in the genesis phase, leading to the
vanishing quadratic action for tensor and scalar fluctuations
in the $t\to-\infty$ limit.
This implies that the validity of the perturbative expansion is questionable
early in the genesis phase, which is, in fact, worse than what is required
for violating the no-go theorem, i.e., ${\cal F}_T\to 0$ as $t\to-\infty$.

\section{Summary}

In this paper, we have generalized the no-go argument of Ref.~\cite{Libanov:2016kfc}
to the full Horndeski theory and shown that non-singular cosmological models
with flat spatial sections are in general plagued with gradient instabilities
or some pathological behavior of tensor perturbations.
We have presented an explicit example which is free from singularities and
instabilities, but has a vanishing quadratic action for the tensor perturbations
(and for the curvature perturbation as well) in the asymptotic past.
To circumvent the no-go theorem,
it is therefore necessary to go beyond the Horndeski theory.
One direction is to consider a (yet unknown) multi-field extension of the Horndeski
theory~\cite{Kobayashi:2013ina,Damour:1992we,Deffayet:2010zh,Padilla:2012dx,Ohashi:2015fma,Horbatsch:2015bua}
in which scalar fields are coupled non-trivially to each other.
Another is extending further the single-field Horndeski theory
as has been done recently e.g. in
Refs.~\cite{Gleyzes:2014dya,Gao:2014soa,Zumalacarregui:2013pma,Domenech:2015tca,Crisostomi:2016tcp,Crisostomi:2016czh,Achour:2016rkg,Ezquiaga:2016nqo}.
It would be interesting to explore to what extent the no-go argument for non-singular cosmologies can be generalized.

%--- Acknowledgements ---%--- Acknowledgements ---%--- Acknowledgements ---%
\acknowledgments
This work was supported in part by the JSPS Grants-in-Aid for
Scientific Research No.~16H01102 and No.~16K17707.
%--- Acknowledgements ---%--- Acknowledgements ---%--- Acknowledgements ---%

%-------------------------------------------------------------------%

%\appendix

%-------------------------------------------------------------------%

%---------   References   ---------%

\end{document}